\documentclass[conference]{IEEEtran}
\usepackage[normalem]{ulem}
\usepackage[hyphens]{url}
\usepackage[noadjust]{cite}
\usepackage{amsmath,amssymb,amsfonts}
\usepackage{algorithmic}
\usepackage{graphicx}
\usepackage{textcomp}
\usepackage{xcolor}
\usepackage{multicol}
\usepackage{multirow}
\usepackage{tikz}
\usepackage{amsmath}
\newcommand{\ignore}[1]{}
\usepackage{tabularx}

\begin{document}
\bstctlcite{IEEEexample:BSTcontrol}

\title{\Large \bf Agon: A Scalable Competitive Scheduler for Large Heterogeneous Systems}

\author{
    \IEEEauthorblockN{Andreas Prodromou}
    \IEEEauthorblockA{
        \textit{NVIDIA}\\
        prodromou.andreas@gmail.com
    }
    \and
    \IEEEauthorblockN{Ashish Venkat}
    \IEEEauthorblockA{
        \textit{University of Virginia}\\
        venkat@virginia.edu
    }
    \and
    \IEEEauthorblockN{Dean M. Tullsen}
    \IEEEauthorblockA{
        \textit{University of California, San Diego}\\
        tullsen@cs.ucsd.edu
    }
}
\maketitle

\begin{abstract}
This work proposes a competitive scheduling approach, designed to scale to large heterogeneous multicore systems. 
This scheduler overcomes the challenges of (1) the high computation overhead of near-optimal schedulers,
and (2) the error introduced by inaccurate performance predictions.  This paper presents Agon, a neural network-based classifier that selects from a range of schedulers, from simple to very accurate, and learns which scheduler provides the right balance of accuracy and overhead for each scheduling interval.  Agon also employs a de-noising frontend allowing the individual schedulers to be tolerant towards noise in performance predictions, producing better overall schedules.  By avoiding expensive scheduling overheads, Agon improves average system performance by 6\% on average, approaching the performance of an oracular scheduler (99.1\% of oracle performance).

\end{abstract}

\section{Introduction}
\label{sec:introduction}

In recent years, there has been an increasing interest in heterogeneous hardware, both in industry and in the research literature.  Processors or systems that allow diverse execution engines maximize the likelihood that any particular thread or application finds an execution core most suited to its execution, maximizing performance and energy efficiency.  This
execution diversity can take the form of application-specific accelerators~\cite{buchty2012survey,putnam2015reconfigurable,jouppi2017datacenter}, as well as general purpose
cores with diverse micro-architectures~\cite{conf/micro/2003/kumar,conf/isca/2004/kumar} or even Instruction Set Architectures~\cite{conf/asplos/2012/DeVuyst, conf/isca/2014/Venkat,conf/hpca/2019/Venkat,conf/asplos/2016/Venkat,barbalace-popcorn-1,barbalace-popcorn-2}.
\ignore{
due to a multitude of benefits that accompany it. The coexistence of various micro-architectures, Instruction Set Architectures (ISAs), and application-specific accelerators, provides among others increased performance at lower power consumption. 
}

Harvesting the full potential of heterogeneity is not trivial, as it requires effectively assigning diverse tasks with constantly changing resource needs, to the best combination of hardware engines
at any point in time.  The larger the system and the greater the level of diversity, the more difficult it is to find the best solution.
A large body of work can be found in the literature with proposals to address resource management in heterogeneous hardware~\cite{conf/asplos/2014/SomuMuthukaruppan,conf/isca/2012/Craeynest,het-datacenters-heuristics-preselection}. 

A variety of proposals~\cite{conf/isca/2012/Craeynest,chen2007mapping,conf/dac/2016/zheng, baldini2014predicting} focus on predicting the performance of each workload on each core of 
a heterogeneous system. This information can then be passed to the scheduler which then uses it to generate a schedule (workload-to-core assignment). While several proposals have been successful in predicting runtime characteristics with high accuracy, many just assume the presence of a scheduling algorithm that can utilize their predictions.  With small or less heterogeneous systems this assignment can be trivial, assuming the predictions are accurate.  As systems get larger and more diverse, however, the workload-to-core assignment becomes computationally prohibitive, with the cost of the assignment algorithm potentially overwhelming the gains from a good schedule.
\ignore{real scheduler, however, needs to consider the impact of placing each workload on each core, resulting in algorithms with high complexity. These algorithms have scalability issues as the number of cores rises.
}

Thus, we find that in a complex system, a scheduler can fail (or perform suboptimally) in two ways -- it can identify a poor schedule, or it can spend so much time computing the schedule that it significantly defers useful computation.
\ignore{
We find that due to various reasons a scheduler in a realistic scenario can operate sub-optimally regardless of its complexity and accuracy. Furthermore, prior work demonstrates that the underlying system itself can reduce the difficulty of the scheduling problem enough to avoid the use of expensive, guaranteed-optimal-solution algorithms. 
}

The fact that scheduling needs can differ between cases is the foundation of this work. We propose {\em Agon} -- a scalable competitive scheduler that selects the appropriate scheduling algorithm (from a number of options), given the state of the system, such that it maximizes overall system performance.  For example, in cases where there is significant gain between the best schedule and a ``good'' schedule, it might be profitable to employ a high-cost scheduler, but in another case where many of the potential schedules are near-optimal it may simply suffice to employ a low-cost scheduler.
\ignore{
For example, assume a case of the scheduling problem, where most possible schedules lead to optimal performance. In such a case, there would be no need for an advanced, high-overhead scheduler. We could instead use a more efficient heuristic that does not guarantee an optimal solution. Agon is designed to be the arbitrator and it is orthogonal to the choice of schedulers and performance predictors a system's designer might choose.}
Thus, Agon is not a scheduler, but an arbiter between schedulers.

We design Agon as a trainable neural network module, and further explore and evaluate multiple configurations and neural network architectures that can be used -- some focused on simplicity and efficiency, while others focused on improving Agon's prediction accuracy. Agon has a dual-part architecture. Its novel front-end is trained to filter out performance prediction error (denoiser), much like convolutional neural networks that remove noise from images. Its back-end is a typical classifier that receives performance predictions and outputs a choice of scheduling algorithm. 

This work makes the following major contributions.
\begin{itemize}
    \item We design Agon, an arbiter of schedulers, that is capable of  selecting a competitive and a performance-optimal workload-to-core assignment, while incurring an acceptable scheduling overhead in terms of its overall computational complexity.
    \item We perform extensive evaluation of Agon on a variety of underlying heterogeneous systems, multiple datasets, and multiple scheduling goals.
    \item Our trained neural network module is capable of sustaining noises in performance prediction errors, and generates an optimal schedule for each of our problem instance with an accuracy of 89\%.
    \item We find that simply due to avoiding scheduling overhead when possible, average system performance improves by $\sim6\%$ compared to a system equipped with the best scheduler available, while at the same time we approach the performance of an oracular scheduler selector (99.1\% of oracle performance).
\end{itemize}

We first present an overview of related work in Section \ref{sec:related}. We then motivate our work via a detailed exploration of all the variables that can affect which scheduler should be used in Section \ref{sec:motivation}. We discuss the details of Agon's architecture (Section \ref{sec:arch}), as well as the details of our dataset (Section \ref{sec:dataset}) and evaluation methodology (Section \ref{sec:methodology}). Section \ref{sec:results} present our results and finally Section \ref{sec:conclusion} concludes the paper.
\section{Background and Related Work}
\label{sec:related}
Kumar, et al introduced the single-ISA heterogeneous architectures~\cite{conf/micro/2003/kumar} that combine cores with different microarchitectures on the same chip. Multiple processor vendors currently offer products that feature this design~\cite{big-little,cell,nvidia-kal-el}. \textit{General-purpose} heterogeneous-ISA multicore architectures \cite{conf/asplos/2012/DeVuyst,conf/isca/2014/Venkat,conf/hpca/2019/Venkat,barbalace-popcorn-1,barbalace-popcorn-2}, include microarchitecturally heterogeneous cores that implement different ISAs and demonstrate significant added benefits compared to its single-ISA counterpart. A recent proposal by Venkat, et al.~\cite{conf/hpca/2019/Venkat} further demonstrates how a heterogeneous-ISA CMP can be based on feature-diverse variations of a single ISA, significantly alleviating several barriers to adoption.  In this work, we choose to evaluate our mechanisms on heterogeneous-ISA multicore architectures, primarily because such architectures significantly increase the difficulty of the scheduling problem.  We use the three ISAs, Thumb~\cite{thumb}, Alpha\cite{alpha}, and x86-64\cite{intelx86}, employed by Venkat and Tullsen~\cite{conf/isca/2014/Venkat} in their design space exploration of heterogeneous-ISA architectures. Our proposal is however orthogonal to the underlying system and can be (re-trained and) used with any combination of cores, systems, and performance predictors.

Scheduling in the presence of heterogeneous hardware is typically a two-step process. First, a prediction mechanism attempts to estimate the performance/efficiency of each workload on every possible execution option (e.g. heterogeneous cores of the underlying system, independent systems such as heterogeneous servers, etc), resulting in a 2-D matrix of workload-to-core predictions.  This is then fed to the second stage -- the scheduler. The scheduler's task is to receive the prediction matrix and decide on the final job-to-core assignment that optimizes some execution aspect, such as overall performance, bandwidth, and/or power consumption.  However, many state-of-the-art schedulers suffer from scalability issues due to their algorithmic complexity. For example, the Hungarian algorithm~\cite{hungarian} is the only known algorithm besides brute-force that guarantees finding an optimal assignment in polynomial ($O(N^3)$) time. As shown this work, the Hungarian algorithm quickly stops scaling as we apply it on architectures that employ 8 or more cores.

The scalability issues of scheduling algorithms have been studied extensively in the literature in various contexts, such as power management in dynamically heterogeneous multicores~\cite{winter2010scalable, winter2008scheduling, bower2008impact}, and real-time systems~\cite{brandenburg2008scalability}.  
In particular, Winter, et al.~\cite{winter2010scalable} assume the decay of CMPs throughout their lifetime due to permanent errors, which lead to dynamic heterogeneity and evaluate how each scheduling algorithm performs. They also propose an iterative scheduling solution termed ``local'' scheduling.  However, unlike prior work, our goal is not to design a new scheduling heuristic, but to assume the existence of multiple scheduling algorithms and build a mechanism that selects the appropriate algorithm to be used.

Numerous proposals from the literature focus primarily on the first step of scheduling, i.e., workload-core performance prediction. Some simply assume the existence of a fast, optimal scheduler and focus entirely on improving prediction accuracy~\cite{prodromou2019deciphering, prodromou2019samos,conf/dac/2016/zheng,shelepov2009hass}, while others also attempt to alleviate its overhead. Two methods are commonly utilized to achieve that -- (a) iterative methods perform job migrations upon scheduling intervals, slowly progressing towards a better schedule but likely sampling poor schedules along the way~\cite{shelepov2009hass, liu2013dynamic, teodorescu2008variation}, and (b) divide-and-conquer methods that logically divide large systems into smaller independent scheduling islands~\cite{chen2009efficient}, allowing the use of more complex algorithms operating with limited-scope knowledge. Scheduling islands are a reasonable solution, especially for very large systems. However, we show that, in many instances, limited knowledge can have a significant impact on overall system performance.

The Agon scheduler we propose is a classification neural network~\cite{benediktsson1990neural}, with each class corresponding to the scheduling algorithm to be used.
Agon explores various neural network architectures, such as basic Feed Forward Networks (FFN), convolutional layers, and convolutional autoencoder architectures used for denoising their input~\cite{vincent2008extracting,gondara2016medical}. Convolutions and convolutional autoencoders are typically used with image inputs. In this work, we are able to utilize these architectures and operate successfully on 2-D performance prediction matrices. In fact, we find that denoising autoencoders can remove performance prediction errors almost entirely when properly trained.

\section{Motivation}
\label{sec:motivation}
\vspace*{-0.05in}
This work is motivated primarily by the fact that determining optimal or near-optimal schedules
for large, heterogeneous systems is difficult to get right and computationally expensive.
In this section, we discuss the factors that make the scheduling problem hard -- (a) the computational overhead of the scheduler, (b) the varying difficulty of hardware/workload combinations, (c) the inaccuracy of performance predictions, and (d) scheduling goals.
\subsection{Scheduler Overhead}
The scheduling problem we address in this work is a special case of the linear sum assignment problem~\cite{burkard2009assignment}. The most efficient solution is the Hungarian algorithm, which guarantees to find (one of) the matrix permutations that generates the largest sum on its diagonal, which represents a job-to-core assignment. Many other heuristics have been developed, that may not necessarily guarantee the optimal schedule, but tend to operate at a much lower computational complexity.

The impact of a scheduler's algorithmic complexity has been demonstrated and discussed in prior work~\cite{winter2010scalable} in the context of dynamically heterogeneous architectures. These architectures begin as a homogeneous system, but become heterogeneous due to hardware failures during the system's lifetime. Besides that work however, the scheduling step is often simply assumed in performance prediction proposals. In other words, proposals that design performance predictors, assume the existence of a scheduler that will receive these predictions and generate the actual job-to-core assignment.

In reality, the Hungarian scheduler, with $O(N^3)$ algorithmic complexity, cannot scale to large multicore systems. In fact, we find that in heterogeneous systems with as few as 8 cores, the overhead of Hungarian scheduling outweighs the returned performance benefit. On very large systems (e.g., server-class processors with many cores), this trend is only expected to worsen.

In this work, we measure the scheduler overhead by creating random inputs of varying sizes (4-256 cores) and feeding them to each scheduling algorithm (implemented in Python, running on an Intel i7 processor at 3 GHz). We repeat each experiment 10000 times to account for noise.

\begin{figure}
\centering
  \includegraphics[width=\columnwidth,height=1.8in]{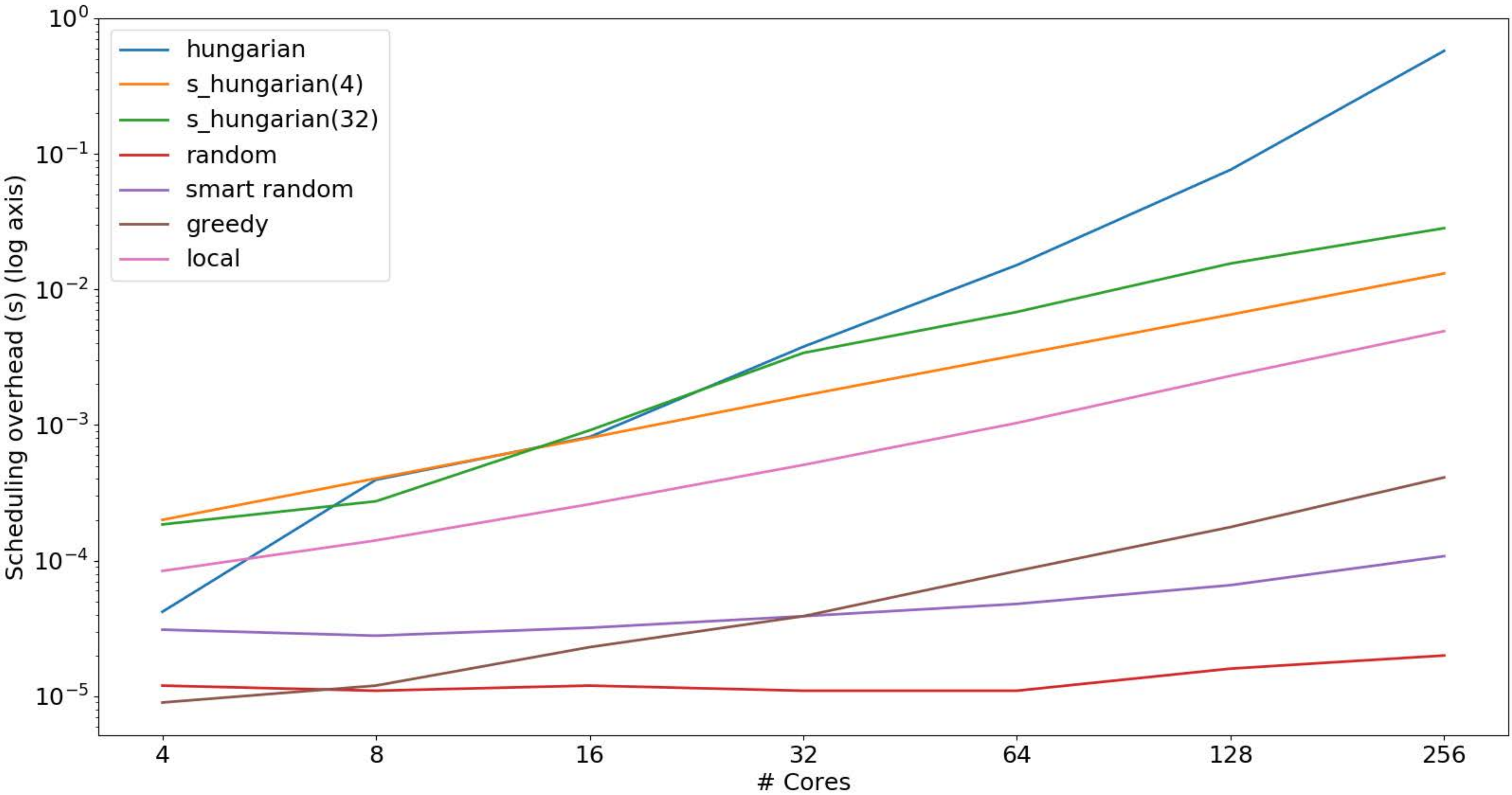}
  \caption{Overhead comparison of 6 scheduling algorithms (Logarithmic Y axis)}
  \label{fig:sched_comp}
\end{figure}

We present our results in Figure~\ref{fig:sched_comp} (note the logarithmic Y axis). We compare six scheduling algorithms. Smart random is the only new scheduler that we develop in this work, while all others are described in prior work. Our measured overheads track each scheduler's algorithmic complexity.  In the remainder of this section, we will show that higher scheduler overhead does not necessarily lead to better performance, and that the returned performance may not necessarily be enough to outweigh the scheduling cost.

In this work we compare the following schedulers:

\textbf{Hungarian Algorithm:} Guarantees finding one (of many in case of ties) optimal solution to the linear assignment problem in $O(N^3)$ (Jonker et al.~\cite{hungarian}).

\textbf{Serial Hungarian:} System is divided into smaller pods. In our exploration, we evaluate pod sizes of 4 and 32. Each pod internally runs the Hungarian algorithm and pods run serially. Algorithmic complexity remains $O(N^3)$, but N is the pod size.

\textbf{Random:} Given N workloads and assuming their order corresponds to the core each one executes, the random scheduler returns a shuffled list with algorithmic complexity $O(N)$.

\textbf{Smart Random:} If possible, this scheduler avoids assigning floating point workloads on the low-power Thumb cores (which do not have FP hardware support). Besides this one rule, all other decisions are random. This scheduler has $O(N)$ complexity.

\textbf{Local:} Given some starting schedule (random at first), this algorithm randomly partitions the schedule into two halves and swaps workloads pairwise. If the new assignment is not beneficial compared to the previous one, swaps are reversed. This process is repeated N/2 times and spans N/2 consecutive scheduling intervals. It can be implemented with $O(N^2)$ complexity. 
\begin{figure}
  \includegraphics[width=\columnwidth,height=1.8in]{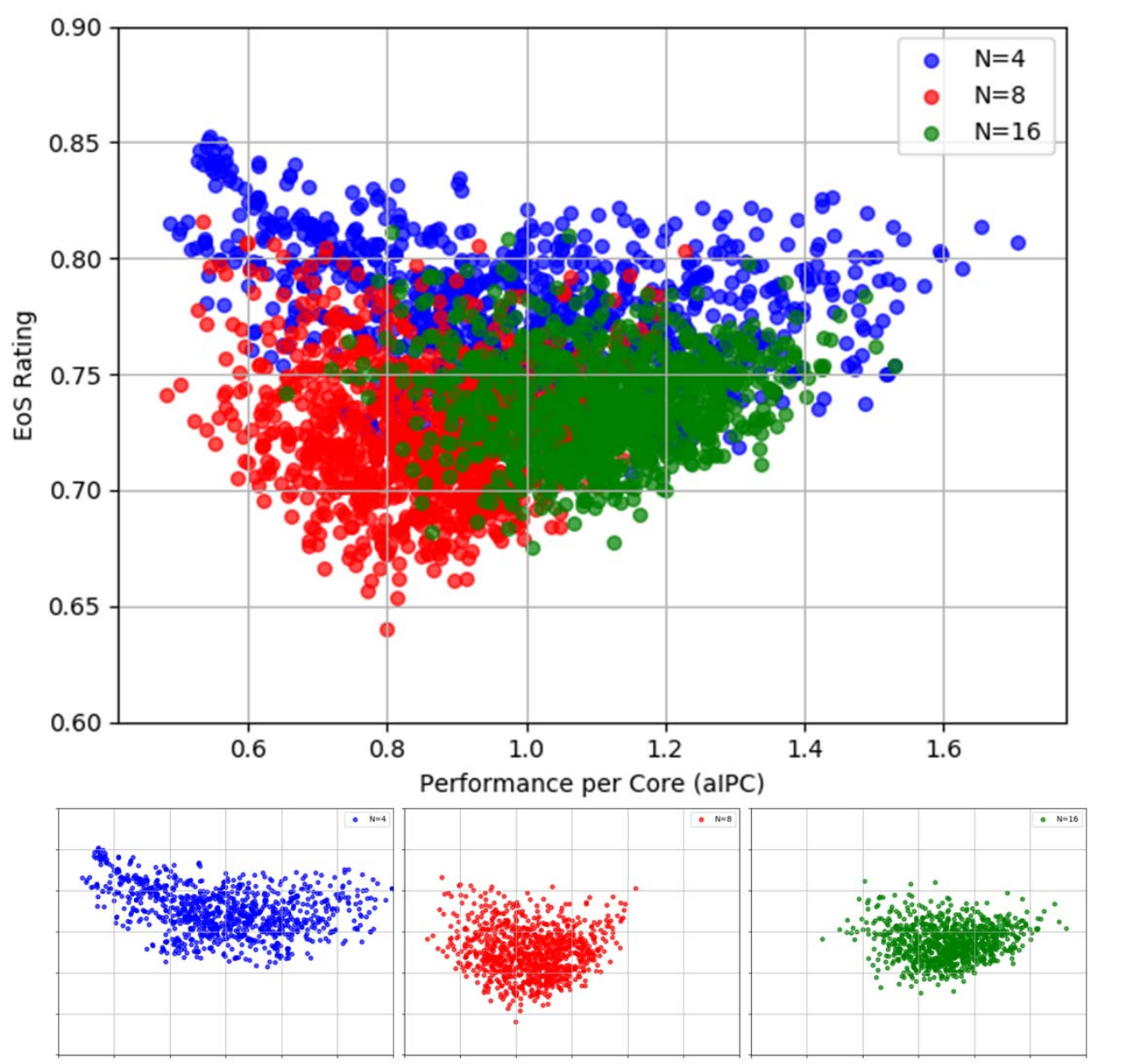}
  \caption{Modified Ease of Scheduling (EoS) Evaluation}
  \label{fig:eos}
\end{figure}

\textbf{Greedy:} Sorts workloads according to ILP and cores according to performance. Final schedule is a one-to-one assignment between the two lists. Can be implemented with $O(N\log{}N)$ complexity. 
\subsection{HW/SW heterogeneity and EoS}
\label{ssec:eos}

The first factor that can affect scheduler quality in our list is core heterogeneity.  Rather than limit our study to a single instance of a heterogeneous system, we examine a large number of combinations taken from 600 unique heterogeneous cores and 3 ISAs, which is the extent of our dataset. Given a randomly selected N-core system, its member cores might be closer to each other in terms of performance or the difference could be quite significant. Typically, as inter-core variance increases, the difficulty of scheduling also increases.

Software heterogeneity affects scheduling in a similar fashion as hardware heterogeneity.  Software can be compute- or memory-bound, can include floating-point or SIMD operations, can have varied register pressure, etc. Prior work has demonstrated that software, and even distinct phases within each program present ``ISA affinity''~\cite{conf/isca/2014/Venkat} -- an inherent preference to one ISA over others. This can be due to obvious factors such as the presence or lack (e.g. Thumb) of FP hardware, or more subtle factors such as register pressure, addressing modes supported, code density, etc.

\begin{figure}
  \includegraphics[width=\columnwidth,height=1.4in]{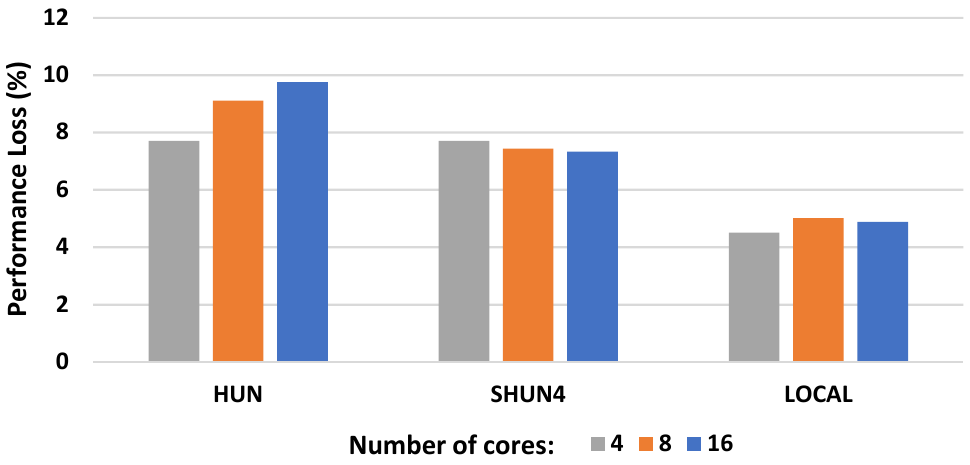}
  \caption{Impact of prediction error on three schedulers (Hungarian, serial hungarian with pod size=4, local). Predictor accuracy is modeled with a bell curve (mean error=0.22, standard deviation=0.35).}
  \label{fig:pred_error_impact}
\end{figure}

Combined, hardware and software heterogeneity can significantly impact a scheduler's ability to find a good job assignment. Prior work proposed an Ease-of-Scheduling (EoS) metric that quantifies the impact of these factors in respect to the scheduler's outcome in a single number~\cite{prodromou2019deciphering}. 
Figure~\ref{fig:eos} shows that for each system size (4/8/16 cores) scheduling difficulty can vary significantly, with some systems being quite easy to schedule, even for a random scheduler, while others can be quite challenging. Subfigures separate each system size and provide a non-overlapped view. In accordance to prior work, we use our EoS results to define the benchmark systems of this study. For each N value (4/8/16), we select the easiest (highest dot), hardest (lowest), lowest-performing (leftmost), and highest-performing (rightmost). In the remainder of this paper, benchmark systems are labeled as \textit{E, H, LP, or HP} respectively, followed by the number of cores in the system. For example, \textit{H4} is the label for the hardest-to-schedule 4-core system, and \textit{LP8} is the lowest-performing, 8-core system.

\subsection{Prediction Error}
Agon assumes that some predictive unit generates the scheduler's input. As such, we can safely assume that the scheduler, regardless of its complexity, will be faced with erroneous inputs. The quality of the predictor can significantly affect scheduling quality. 

We measure the impact of prediction error on three schedulers: Hungarian, serial Hungarian, and local. Random and greedy schedulers do not depend on performance prediction and are immune to prediction error. For fair evaluation, we assume the existence of a good predictor, since as prediction quality drops, all schedulers tend to converge to the performance of a random scheduler. Prior work has evaluated numerous such predictors, using a variety of different machine learning algorithms. We choose a Decision-Tree based predictor~\cite{prodromou2019deciphering,prodromou2019samos} and assume it to be the generator of Agon's input matrix, due to its low overhead and high accuracy.

Figure~\ref{fig:pred_error_impact} shows our results. We find that compared to oracular performance predictions, Hungarian algorithms, both the originally proposed as well as the serial version, are affected between $7.2-9.8\%$ on average by the prediction error of a \textbf{good} performance predictor. The local algorithm is affected at a lesser degree ($4.1-4.9\%$).  However, compared to the Hungarian scheduler, local returns up to 21\% and 17\% lower performance on average for oracular and predicted inputs respectively. 

\begin{figure}
  \includegraphics[width=\columnwidth,height=1.4in]{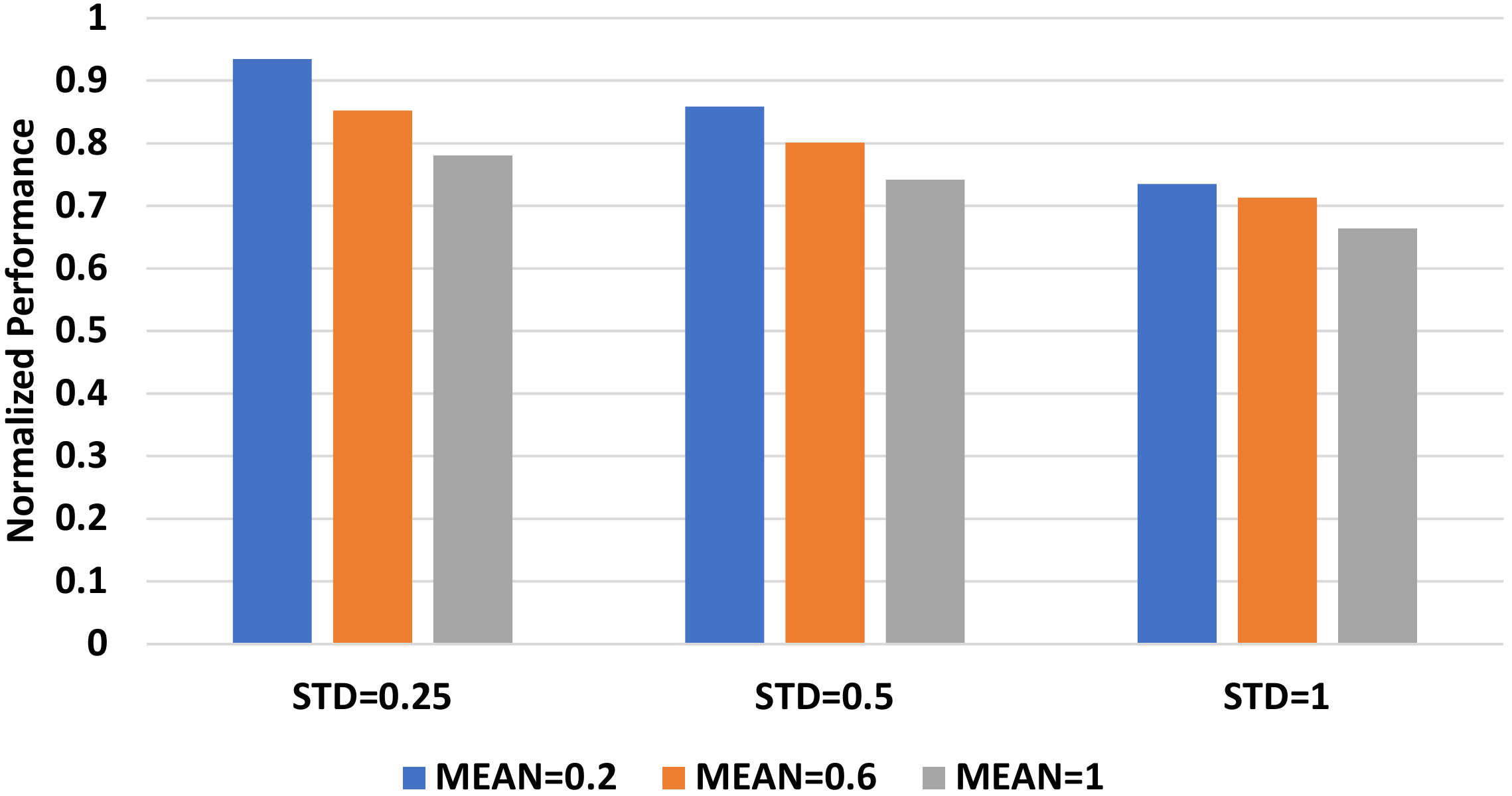}
  \caption{Normalized performance of schedules found using the Hungarian algorithm under performance predictors of varied quality. Predictors are emulated using the mean (MEAN) and standard deviation (STD) of their prediction error.}
  \label{fig:non_slope_impact}

\end{figure}

Figure~\ref{fig:non_slope_impact} presents the impact of prediction error on a single instance of the scheduling problem (i.e., for one particular 8-core system and one set of 8 workloads). Both the system under test, as well as the set of workloads were randomly selected for this experiment. Performance (Y axis) is normalized to the performance of the best possible schedule, which is calculated using the Hungarian algorithm fed with oracular (error-free) performance predictions. All nine schedules presented in the figure are generated using the Hungarian algorithm and each bar corresponds to a different performance predictor that generates the scheduler's input. Each predictor is emulated using the mean and standard deviation of its prediction error. We minimize noise from random decisions by repeating the experiment 10000 times and reporting average performance of schedules. With the worst predictor generating inputs to the scheduler, the derived schedule performs 33\% worse compared to the best possible schedule. We also observe that while mean error in predictions has a significant impact, so does the standard deviation of error our predictor exhibits.
\subsection{Scheduling Goals}
\label{ssec:sched_goals}

In this work, we explore four different scheduling goals -- (a) the \textbf{FIRST100} goal, defines the winner scheduler to be the one that first executes 100 million Alpha instructions (or equivalent work on another ISA), (b) the \textbf{FIRST400} goal selects the scheduler that first executes 400 million Alpha instructions, (c) the \textbf{MOST50} goal selects the scheduler that provides the highest throughput within the next 50 million cycles and, (d) the \textbf{MOST200} goal selects the scheduler that provides the highest throughput within the next 200 million cycles.

In identifying the winner scheduler, we take into account each scheduler's overhead. For example, if the Hungarian scheduler needs 10 million cycles to output an optimal job-to-core assignment and the goal is set to MOST50, this leaves the scheduler with only 40M cycles to execute useful instructions.  Thus, each goal is also impacted by the inherent computational overhead of the scheduler, with the overhead becoming more and more pronounced, as the frequency of scheduling increases.



\section{Architecture}
\label{sec:arch}

Agon is a ``competitive'' scheduler. Conceptually, a number of schedulers (scheduling algorithms) compete to be the one to solve the assignment problem. Agon acts as the arbiter which decides the winner in each instance. In a realistic execution scenario, available input information will be an $N \times N$ performance matrix, representing the (predicted) performance of each of the $N$ in-flight workloads on each of the $N$ cores. A performance predictor is responsible for generating the performance matrix.
However, in our evaluation, we also examine the impact of selecting a scheduler, in the face of oracularly obtained performance matrices (i.e., 100\% accurate predictions).

\begin{figure}
\centering
  \includegraphics[width=0.9\columnwidth,height=1.6in]{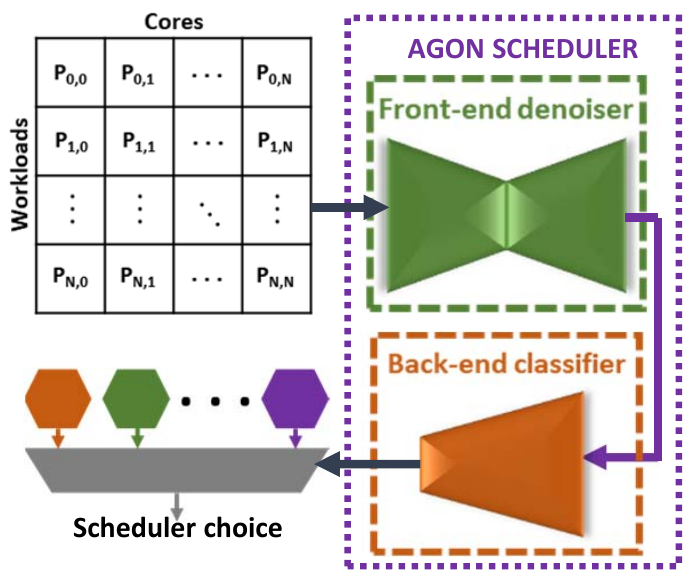}
  \caption{High-Level overview: A performance prediction matrix flows through Agon's front- and back-end modules and finally guides a scheduler selector.}
  \label{fig:high_level_arch}
\end{figure}

Figure~\ref{fig:high_level_arch} shows a high-level overview of our architecture.  At its heart, Agon is a classifier, which receives a performance matrix as input and selects one (of six in this study) available schedulers.  Agon runs the input matrix through its (optional) denoising front-end, followed by its classification back-end . A series of hidden layers in both modules process the information. The classifier's output layer consists of K nodes, with the output nodes using softmax activation~\cite{memisevic2010gated}.  Agon's output is the probability of each scheduler to be the most effective one. The scheduler with the highest probability emerges as the winner.

We explore a variety of possible architectures for Agon, with varied complexity, classification accuracy, and system performance. We describe these architectures in the remainder of this section.
\subsection{Classification Back-End}
This is the main control center of Agon that is responsible for the final decision. We explore two neural network architecture options -- (a) a simple, feed forward architecture (labeled ``deep'' in the remainder of this work), and (b) a convolutional classification architecture (``cnn''). Both architectures result in a softmax output layer with 6 nodes, as we train it for 6-way classification.

As a classifier, the operation of Agon's back-end is quite straightforward. The returned class is the one with the highest (softmax) probability. The input layer is internally ensured to be compatible with the selection of denoiser (described in Section~\ref{ssec:frontend_agon} - via the use of pseudolayers such as ``Flatten'' and ``Reshape''). For example, a convolutional denoiser will output a 2-D matrix, while the simple denoiser will output a flattened 1-D list. Similarly, a feed-forward classifier will expect a flat input, while a convolutional one will expect a 2-D input.

\begin{table}
\begin{center}
\begin{tabular}{|cccc|}
\hline
\textbf{Layer ID} & \textbf{Layer Type} & \textbf{\# Nodes} & \textbf{Activation} \\ \hline
1 & Dense & 128 & ReLU \\ 
2 & Dense & 32 & ReLU \\ 
3 & Dense & 6 & softmax \\ \hline
\end{tabular}
\end{center}
\caption{Layer breakdown: Deep FFN classifier}
\label{tab:deep_classifier}
\end{table}

\begin{table}
\begin{center}
\begin{tabular}{|ccc|}
\hline
\multicolumn{3}{|c|}{\textbf{CNN section of classifier}} \\
\hline
\textbf{Layer Type} & \textbf{\# Filters} & \textbf{Kernel Size} \\ \hline
Conv2D & 64 & (4,4) \\
MaxPooling2D & NA & (2,2) \\
Flatten & \multicolumn{2}{c|}{\textit{Pseudolayer: Flattens output matrix }} \\    
\hline
\hline
\multicolumn{3}{|c|}{\textbf{FFN section of classifier}} \\
\hline
\textbf{Layer Type} & \textbf{\# Nodes} & \textbf{Activation} \\ \hline
Dense & 64 & ReLU \\ 
Dense & 32 & ReLU \\ 
Dense & 6 & softmax \\ \hline
\end{tabular}
\end{center}
\caption{Layer breakdown: Deep convolutional classifier}
\label{tab:conv_classifier}
\end{table}
Tables \ref{tab:deep_classifier} and \ref{tab:conv_classifier} present more details about the architectures we explore. We explore various architectural options and select the two most efficient for our study.

\subsection{Denoising Front-End}
\label{ssec:frontend_agon}

We propose an autoencoder-based front-end, responsible for removing prediction error (if any) from the input matrix, which is produced by a performance predictor. This idea is inspired by neural-network-based noise-reduction techniques for images. Given the success of autoencoders in denoising images, we expect it work well in our example. To the best of our knowledge, this is the first neural network-based proposal that attempts to reduce prediction error after predictions are made.

Denoising autoencoders receive some realistic input during training (e.g., a noisy image), and are trained to output the oracular (i.e., image without noise) data at their outputs.  We adapt this methodology and train our denoising frontends with realistic, non-oracular predictions from our performance predictor module, and use our oracular dataset as labels. Our datasets are described in more detail in Section~\ref{sec:dataset}.

Similarly to the back-end classifier, we explore two denoiser architectures for Agon -- (a) a simple, feed forward architecture, and (b) a convolutional architecture. The feed forward architecture, labeled ``simple'' for the remainder of this work, consists of three layers. For 8-core systems, these layers have 64 nodes at the input and output layers, and 32 nodes in the one hidden layer. While our evaluation includes deeper architectures with more hidden layers, this simple 3-layer network provides the best outcomes in most of our experiments.

Convolutional neural network architectures are typically used along with images, due to their ability to recognize shapes, forms, and other elements found in images (information that is typically lost when the 2D image is flattened into a 1D array of pixel values). Convolutional autoencoders are particularly successful at reducing noise from images. 
A (predicted) performance matrix might not have any meaning as an image. However, we claim that a convolutional ``deconstruction'' can still be meaningful.  Assume we have a system of 8 cores ($8\times8$ input matrix) and a convolutional $4\times4$ window that \textit{slides in 2 dimensions} over the input. Also assume that the input matrix is a greyscale image (higher values are darker). The (sliding) convolutional window will look at the input in groups of 4x4. Each 4x4 sub-matrix might include a shape -- for example a dark diagonal, a straight line, a shape similar to a block in the game of Tetris, a diamond, etc. All these ``shapes'' are meaningful to the denoiser and through training over many such subfigures, it can learn to darken or whiten (increase or decrease the predicted performance) certain entries of these sub-matrices such that the shape is consistent with its training.  Because of the 4x4 sliding window, each entry of the matrix is examined 16 times, in 16 different neighboring sum-matrices, allowing the denoiser to have even higher confidence in its changes.


\begin{table}
\begin{center}
\begin{tabular}{|cccc|}
\hline
\textbf{Layer ID} & \textbf{Layer Type} & \textbf{\# Nodes} & \textbf{Activation} \\ \hline
1 & Dense & 32 & ReLU \\ 
2 & Dense & 64 & ReLU \\ \hline
\end{tabular}
\end{center}
\caption{Layer breakdown: Simple FFN denoiser}
\label{tab:simple_denoiser}
\end{table}

\begin{table}
\begin{center}
\begin{tabular}{|ccc|l|}
\hline
\textbf{Layer Type} & \textbf{\# Filters} & \textbf{Kernel Size} & \\ \hline
Conv2D & 32 & (3,3)  & \multirow{4}{*}{\rotatebox[origin=c]{90}{Encoder}} \\
MaxPooling2D & NA & (2,2) & \\
Conv2D & 32 & (3,3) & \\
MaxPooling2D & NA & (2,2) & \\ \hline\hline
Conv2D & 32 & (3,3) & \multirow{4}{*}{\rotatebox[origin=c]{90}{Decoder}} \\
UpSampling2D & NA & (2,2) & \\
Conv2D & 32 & (3,3) & \\
UpSampling2D & NA & (2,2) & \\
Conv2D & 1 & (3,3) & \\ \hline
\end{tabular}
\end{center}
\caption{Layer breakdown: Deep convolutional denoiser. All layers use the ReLU activation function. Where it applies, padding is set to ``same''}
\label{tab:conv_denoiser}
\end{table}




Besides the two denoiser architectures, we also evaluate Agon without a denoiser. Tables \ref{tab:simple_denoiser} and \ref{tab:conv_denoiser} present more details for our proposed architectures.  We use Tensorflow's Keras API and design each classification module with a non-trainable input layer, to create standalone front- and back-end modules that can be easily combined.

\subsection{Front-End, Back-End Synergy}
Agon's two-part decoupled architecture generates some opportunities for synergy between the front- and back-ends. An individually trained front-end denoiser with high efficiency can be used in conjunction with a classifier that has been (also individually) trained using \textit{oracular} data. Assuming successful denoising, there is no reason to train a classifier with data that comes straight from a predictor. On the other hand, if the classifier expects stellar, error-free predictions, it could fail in the presence of even a small error, if any, is left from the denoiser.

We can also train the denoiser and the classifier together as one linked unit, in which case, we observe an interesting behavior -- the front-end no longer acts as a denoiser, and instead, it learns to transform the erroneous input into a new matrix (which may or may not be error-free, may or may not look similar to the actual input), but generated with the back-end in mind. In other words, the front-end \textit{generates a classifier input from the real input, such that the classifier's accuracy increases.}  Our evaluation shows that this indeed happens, with linked Agon models outperforming even classifiers trained using oracular data.
\section{Dataset Generation}
\label{sec:dataset}
This section describes our methodology for gathering and generating data needed in our study. 

We start with our cycle-accurate simulation-based dataset of 72 workloads, each simulated in isolation on 600 different cores. These cores are divided into three groups of 200 (heterogeneous) cores each, one group for each of the three ISAs used in our study: Arm's Thumb, Alpha, and Intel's x86-64. The workloads are benchmark phases of 100 million \textit{Alpha} instructions each, identified using Simpoint, from the SPEC 2006 benchmark suite. Our LLVM-based compiler allows us to map code regions (phases) to their equivalent code regions in compiled executables for a different ISA.  The dataset contains workload profiling results (Gem5, McPat), as well as microarchitectural descriptions of each core. Overall, this dataset contains 43,200 entries. Each entry contains 30 features that characterize the workload (e.g., cache access statistics, fetch/issue rates, performance in aIPC (Alpha-IPC)), and 18 features that describe the core's architecture (e.g., in-order vs. out-of-order, cache hierarchy, SIMD support, number of functional units, registers).
\subsection{Generation of Agon's dataset}
Agon requires a different dataset for its training, as its input is an $N \times N$ matrix of predicted performance for each workload-core pair (N is the number of cores of the multicore system under test).

We first generate a list of $3 \times 1000$ multicore systems, 1000 systems for each N (4/8/16). 
The cores of each system are chosen randomly out of the 600 cores in our dataset ($1.6\times10^{22}$ possible 8-core system combinations). We force our system generator to ensure each system includes at least one core from each of the three ISAs, in order to ensure significantly heterogeneous(-ISA) systems that consequently make scheduling more challenging. We then generate a list of 20K workload combinations in a similar way, by randomly picking from our list of 72 workloads ($7.2\times10^{14}$ possible 8-workload combinations). We generate three lists of workload combinations, one for each size of N in our study. All 20K workload selections for each value of N can run on each of the 1000 systems of the same size, leaving us with a dataset of 60M entries (20M for each N).
Each entry (row) in our dataset is populated with the \textit{measured (profiled via simulation)} performances of the workload-core pairs. For example, on 8-core, 8-workload systems, each dataset entry will have 64 performance measurements. This becomes our \textit{oracular} dataset. 

We use the oracular dataset as a comparison point later in our explorations, but it is also necessary for calculating the correct answer (label, scheduler to be used), for each entry of our dataset: Using oracularly obtained information, we execute each scheduler using the same six scheduler  implementations used to generate Figure \ref{fig:sched_comp} and we receive one schedule from each one. For each of the 6 produced schedules 
(which may or may not be different from each other), we calculate its \textit{effective aIPC}, which takes into account the scheduler's overhead. Finally, whichever scheduler produces the highest effective aIPC is considered the correct answer. Even in cases where we don't use an oracular, error-free input matrix, Agon is required to output the labels we calculate using oracular data since, regardless of prediction error, that still is the correct answer.

We generate a total of 4 datasets for this study. The first is oracular which we just described. We also generate two predictor-based datasets, using trained predictors from prior work.  The first is a Decision Tree (DT) predictor, trained with 6/10 benchmarks, while the remaining 4 are reserved for testing as previously-unseen data. We term this ``multiLOGO DT'', due to the multi-LOGO data splitting process used in prior work~\cite{prodromou2019deciphering}.  The second is also a DT predictor, this time however, we train it with the entirety of our dataset to reflect a predictor in a system that executes the same workloads over and over again. In our study, we use it mostly as a comparison point, while we focus on obtaining good results on the multiLOGO-DT. We name this ``fully-trained DT''.

Our last dataset is called the ``statistical'' dataset. Our DT performance predictors'  error can be modeled as a bell curve, with a mean value and standard deviation. Specifically, our multiLOGO DT performance predictor reports $mean=0.25(aIPC)$ and $std=0.35$. We generate the statistical dataset by randomly drawing a prediction error value from the bell-curve that characterizes our predictors and simply applies it on the oracular value we get from the oracular dataset. Statistical error injection makes Agon's job artificially more difficult.  Any real predictor is likely to make errors with some consistency or pattern that Agon could potentially learn.
\subsection{Dataset Imbalance}
\begin{figure}[t]
  \includegraphics[width=0.48\textwidth, height=1.8in]{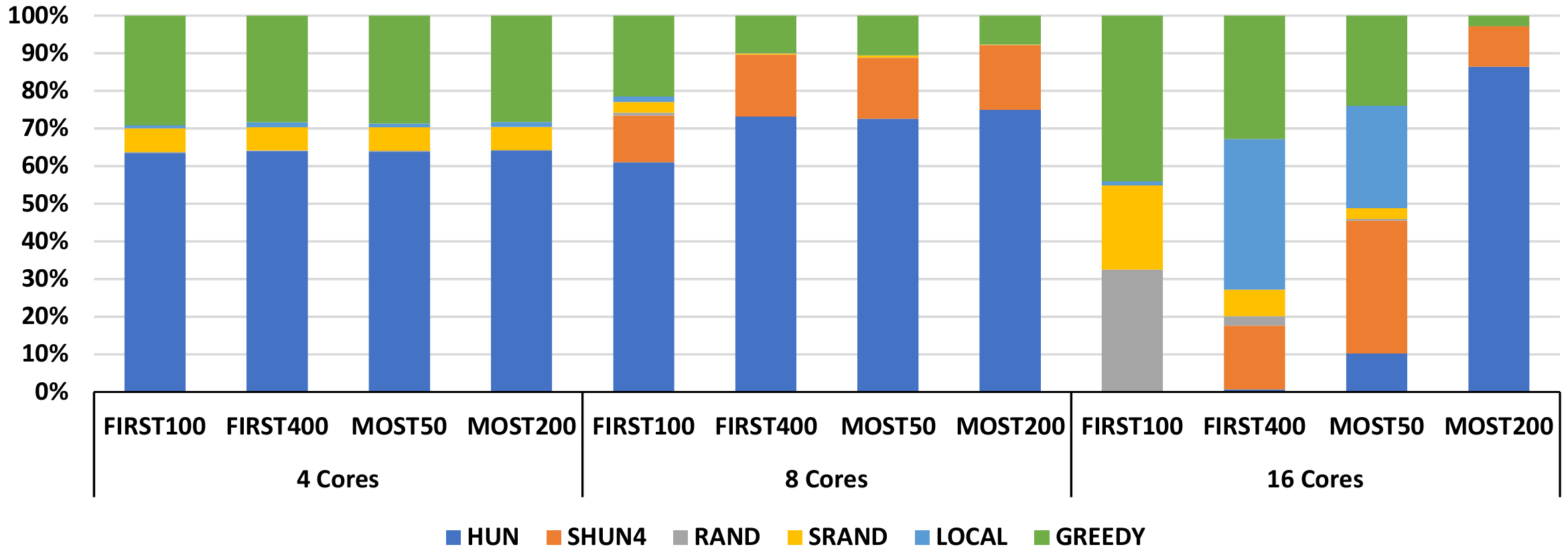}
  \caption{Breakdown of preferred schedulers for each entry in our dataset.}
  \label{fig:sched_pref}
\end{figure}

Figure \ref{fig:sched_pref} presents a breakdown of the labels in our dataset, which correspond to the optimal scheduler choice of each entry in our dataset. The x-axis of the figure presents a variety of \textit{scheduling goals}, as discussed in Section~\ref{ssec:sched_goals}. \textit{HUN} and \textit{SHUN4} are shorthands for the Hungarian and Serial Hungarian (with pod size of 4) schedulers respectively. 


Our results show that our dataset is imbalanced in favor of the Hungarian algorithm, which dominates most bars in the Figure. Dataset imbalance is a challenge in training a good classifier. On a 4-core system for example, the Hungarian scheduler is preferred for the vast majority of cases, due to its low overhead (since we only have 4 cores) and the fact that it tends to result in better performance overall. As systems grow in size, we see a trend to move away from expensive Hungarian-based schedulers and more towards random and greedy (i.e. ultra-low cost) schedulers. However, in systems where we run the scheduler
less frequently (such as \textit{16 cores -- MOST200}), Hungarian again is important, 
because relative overhead is reduced and the importance of a good schedule is increased because it
runs longer.  However, some of our large systems with reduced scheduling frequency, rarely use Hungarian.

This imbalance makes it easy for classifiers to learn to always pick one class. If Hungarian is the correct answer 90\% of the time, always picking it guarantees a 90\% prediction accuracy. This classifier is not useful and could simply be removed from the scheduling process. Dealing with imbalanced datasets is actively pursued in the literature, since many crucial problems require good performance on edge cases (e.g. credit card fraud, self-driving cars). These are also problem categories where mispredicting corner cases can lead to catastrophic errors. A few options found in the literature are upsampling of corner cases, downsampling of popular cases, algorithmic dataset-driven generation of new (random) entries, and using class weights during training. We find upsampling to work best in our study. We clone the entries of under-represented classes until the count of each class in our dataset is roughly even. Due to upsampling, the classifier no longer blindly chooses one option. On the other hand however, we have to be careful when dividing our dataset into training and testing subsets, since we don't want cloned entries to appear in both sets. Statistical error injection is immune to this issue since the applied error will differ between cloned entries.

\section{Evaluation Methodology}
\label{sec:methodology}
We evaluate Agon on 4 datasets, 12 heterogeneous(-ISA) multicore systems, 4 scheduling goals, 6 different schedulers and 6 proposed neural network architectures. This section presents our experimental framework and training methodology.
\subsection{Experimental Framework}
The majority of our framework is developed in Python, using the Tensorflow framework for neural network training and evaluation. We use the Keras API with a Tensorflow backend, due to its simplicity. Our front-end denoisers, back-end classifiers, and complete Agon models are implemented using the \textit{Sequential} model API with the standard Keras layers (Dense, Conv2D, etc).

During our evaluation, we assume that workloads appear in groups of N and each instance is independent. In other words, we assume an 8-core (for example) system that is presented with 8 workloads as soon as the previous batch of 8 workloads finishes. Each batch of 8 workloads represents a new scheduling problem instance (i.e., new predicted performance matrix, new prediction from Agon). In the majority of our results, we report the average performance after all batches finish executing.

We use \textit{effective aIPC} as our performance metric in this work. We use Alpha Instructions Per Cycle (aIPC) to fairly compare the progress of executables of different ISAs, as proposed in prior work~\cite{prodromou2019deciphering}. Effective aIPC represents a system's true performance once scheduling overhead is applied, as discussed earlier in Section~\ref{ssec:sched_goals}. 

\subsection{Agon Training Methodology}
We train a different, independent Agon model for each underlying system. We utilize the \textit{EarlyStopping} callback from Keras to avoid overfitting. This callback automatically stops the training process once accuracy over the validation test ceases to improve for a number of consecutive training epochs. We divide our dataset into three sets: $~80\%$ of our data is assigned to the training set, $~5\%$ for our validation set, and the remaining $~15\%$ forms our test set.

Agon's backend classifier uses ``sparse categorical crossentropy'' as its loss function. 
Agon models that link a denoiser and a classifier for simultaneous training use this loss function as the end goal is accurate classification. When we train front-end denoisers in isolation, we use ``Mean Absolute Error'' as the loss function, since we want the autoencoder's output to be as close as possible to the oracularly obtained performance matrices. Finally, we use the state-of-the-art Adam~\cite{kingma2014adam} optimizer for all our training.

We focus our evaluation on 8-core systems and the ``FIRST100'' scheduling goal to narrow the scope of the work to a size that allows us to come to conclusions without limiting the generality of our proposal. We choose this particular combination due to the healthier distribution of scheduling labels, as shown in Figure~\ref{fig:sched_pref}.

\section{Results}
\label{sec:results}
\subsection{Classifier Accuracy}
\begin{figure}
  \includegraphics[width=\columnwidth,height=1.6in]{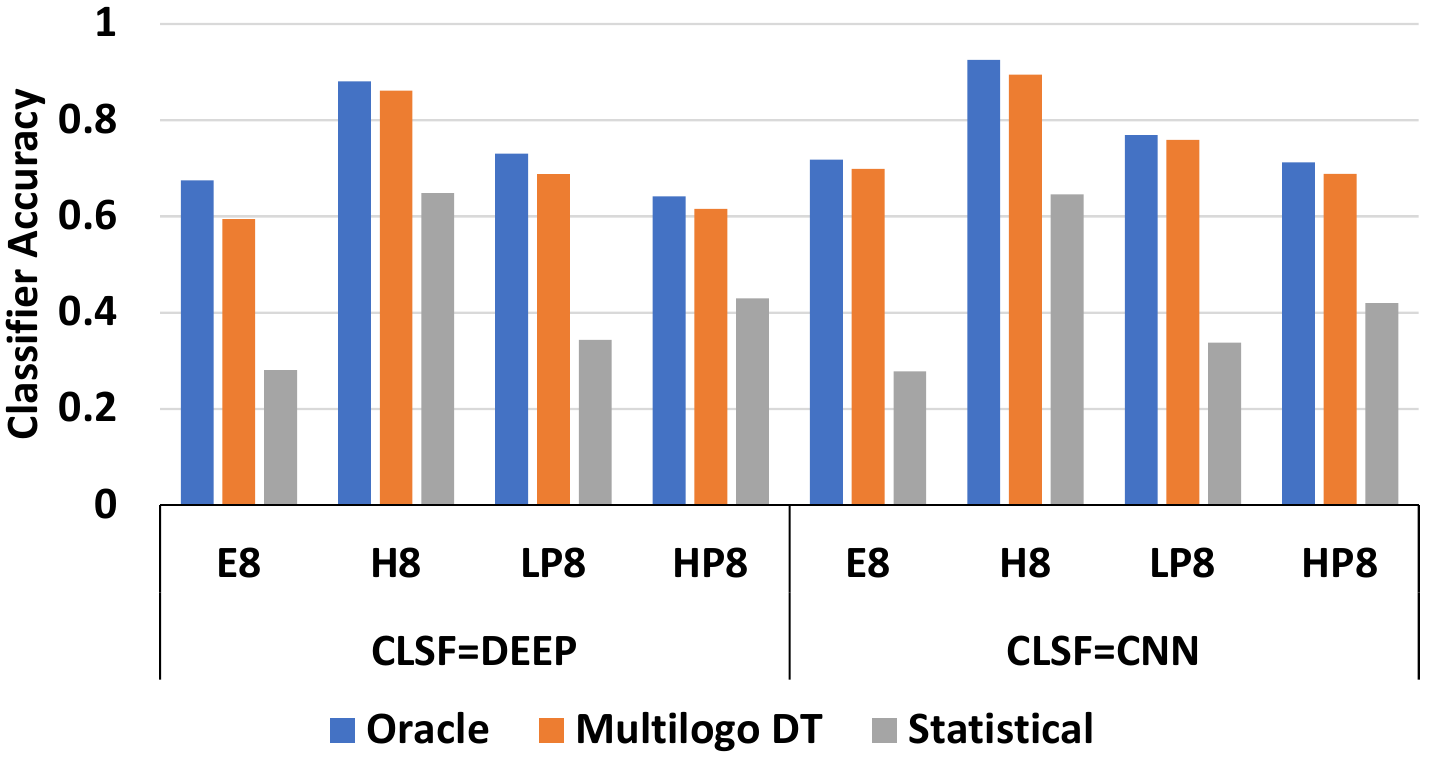}
  \caption{Classifier Prediction Accuracy. DEEP Vs CNN classifier architecture, oracle Vs multilogo Vs statistical datasets.}
  \label{fig:clsf_acc}
\end{figure}

Figure \ref{fig:clsf_acc} presents our classifier's prediction accuracy, for a backend-only configuration (i.e., without a denoiser frontend). Prediction accuracy reveals our classifier's ability to pick the correct scheduling class. In other words, it is the ratio of correct scheduler predictions. For comparison, a random arbiter would have 16.6\% ($1/6$) accuracy. Specifically, we present a two-fold comparison. For each of our 4 benchmark systems we compare the classifier's accuracy under our oracular (unrealistic), multilogo (realistic), and statistical (artificially more challenging) datasets.  We further compare the two classification architectures we propose for Agon. The deep feed-forward architecture is presented on the left half of the figure, while the convolutional classifier architecture is shown on the right half.

\begin{figure}
  \includegraphics[width=\columnwidth,height=1.6in]{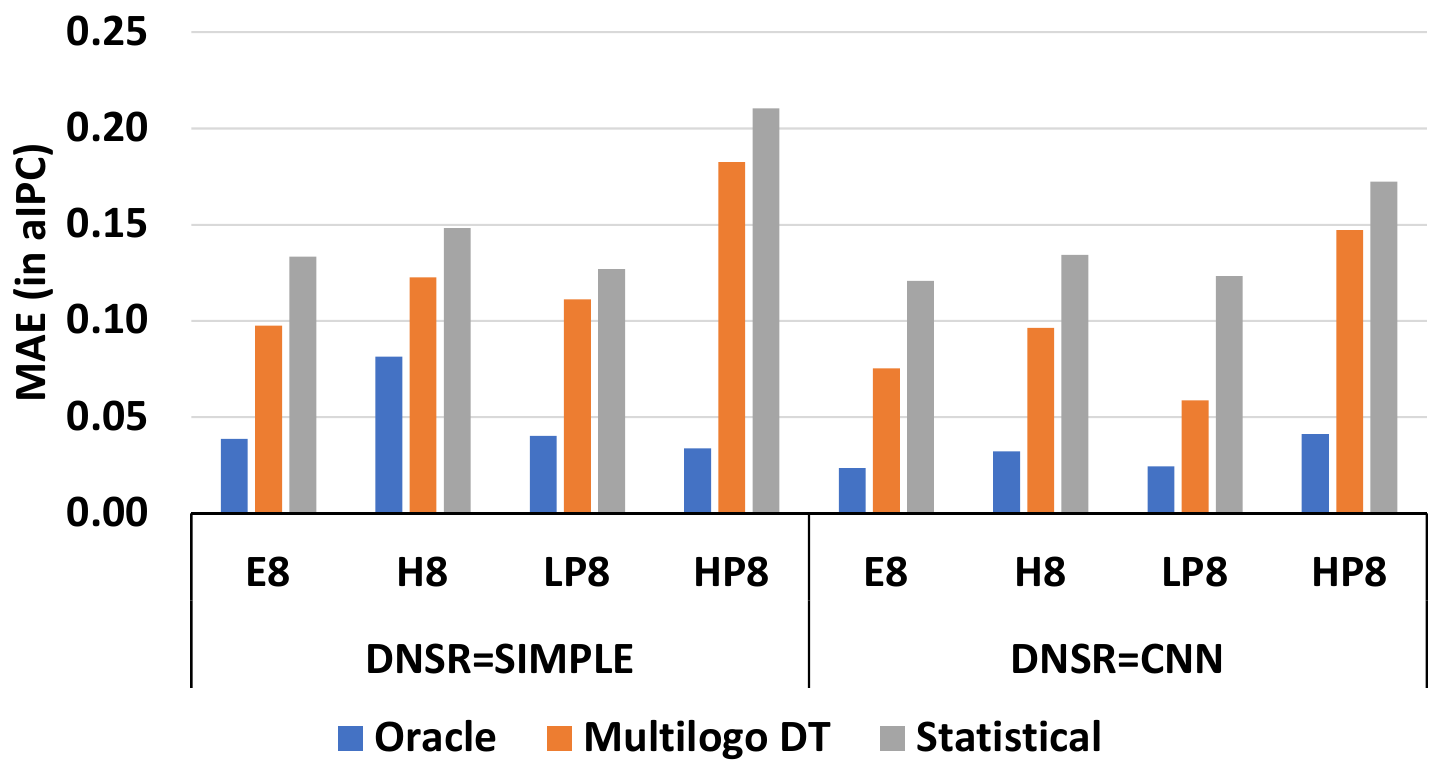}
  \caption{Denoiser loss in MAE (lower is better). SIMPLE Vs CNN classifier architecture, oracle Vs multilogo Vs statistical datasets.}
  \label{fig:dnsr_mae}
\end{figure}

We first observe that accuracy on the realistic dataset is comparable with the unrealistic oracular. Specifically, the difference in prediction accuracy ranges between 8\% (E8, DEEP) in the worst case to 1\% (LP8, CNN), practically within noise levels. This result hints that our classifiers are able to reach the predictability limits of our problem. In other words, the misclassified cases are roughly those that would be misclassified even with oracular information. Due to the good quality of our multilogo predictor and its error consistency, the classifier can still work efficiently with the erroneous input. Results over the statistical dataset show that prediction error inconsistency can severely impact classification quality.

\begin{figure*}[t]
  \includegraphics[width=\textwidth,height=1.6in]{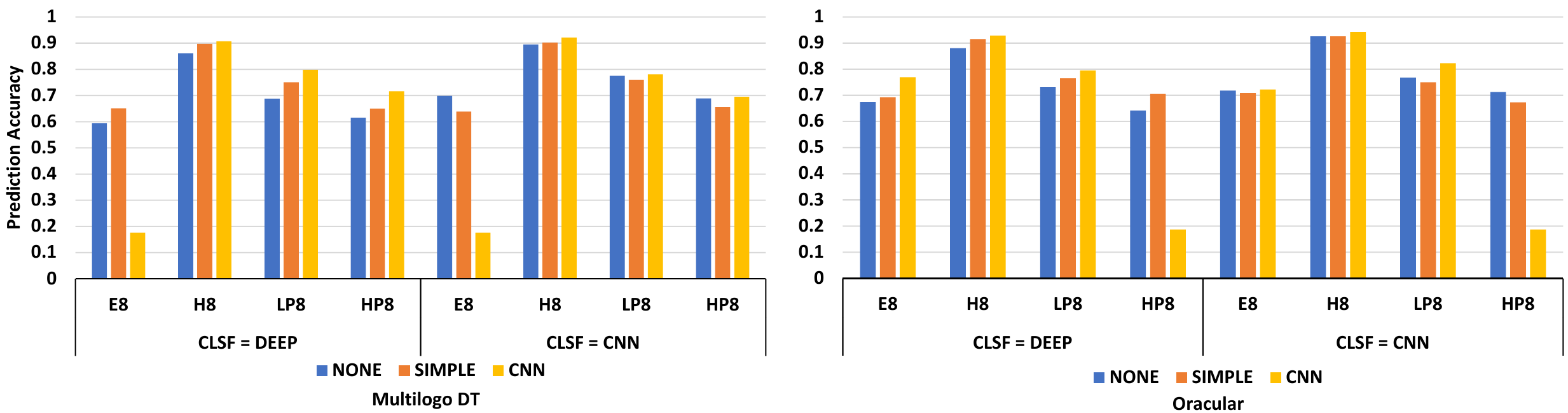}
  \caption{Impact of frontend on Agon's prediction accuracy on multilogo and oracular datasets. Agon's denoiser and classifier are trained together as one unit.}
  \label{fig:dnsr_impact}
\end{figure*}

\begin{figure*}[t]
  \includegraphics[width=\textwidth,height=1.6in]{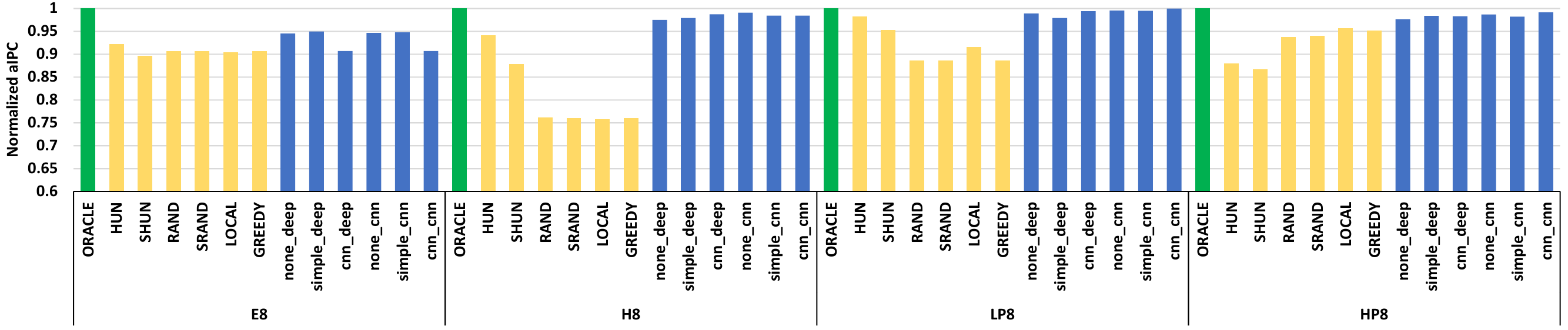}
  \caption{Performance comparison of 12 scheduling strategies, normalized to oracular knowledge of optimal scheduling algorithm for each problem instance. The first 6 strategies are ``always'' strategies, where there is only one scheduler option. Last 6 bars are Agon architectures.}
  \label{fig:agon_perf}
\end{figure*}

Our results further show the convolutional classifier outperforming the simpler deep architecture. Specifically, CNN outperforms DEEP by 3.7-7.1\% (oracular), 3.3-10.4\% (multilogo), across all systems. The two architectures perform almost identically on the statistical dataset, again due to the unpredictability of the input. 

The underlying heterogeneous(-ISA) multicore system is shown to have an impact on predictability as well. In fact, our results show an inverse correlation between a system's EoS (difficulty of a scheduler to find a good schedule), and the system's difficulty level in predicting the most efficient scheduler. Using a convolutional classifier on the multilogo dataset, Agon reports accuracy up to 89\% for the most difficult, lowest-EoS system (H8). The other three systems report accuracy between 68-76\%. The deep architecture reports lower accuracy, but still follows the same trend.

Prediction accuracy is not necessarily representative of the resulting performance we should expect. For example, if we mispredict one problem instance of the E8 system, we should still expect the (wrong -- not most efficient for this instance) scheduler to produce a good schedule, due to the increased odds of success (high EoS). We present a performance comparison later in this section. 
\subsection{Denoiser Accuracy}
Agon's denoiser (when individually trained) attempts to minimize Mean Absolute Error (MAE) and bring the output layer's values as close to ground truth as possible. We present our denoiser MAE results in Figure \ref{fig:dnsr_mae} (lower is better). On its own, this experiment does not provide any insight on how Agon improves with the addition of a denoiser. However, our results demonstrate that the denoising problem is in fact solvable, allowing computer architects to attempt denoising on other problems where hardware predictions are utilized.

Similarly to prior results, we present a comparison between three datasets. Denoising the oracular dataset represents an interesting case, which corresponds to typical autoencoder operation, where the output is trained to match the input, while internally, the dimensionality of the input reduces in the network's hidden layers -- in our ``simple'' architecture, 64 input values reduce to 32 before fanning out to 64 output values. Results on multilogo and statistical, require true denoising, since we ask the frontend module to output oracular values, while receiving erroneous inputs.

In our use case, MAE is measured in aIPC, which in our dataset can be as high as 4.5 aIPC. Agon's frontend successfully restores oracular inputs with absolute error below 0.05 in all cases but one (0.08 MAE), demonstrating that while dimensionality reduces, there remains enough information in the hidden layers to accurately characterize the entire input.

When dealing with multilogo erroneous inputs, denoiser error grows significantly compared to oracular, remaining however between 0.06-0.18 aIPC (0.11 aIPC on average). Finally, the statistical dataset, which is our most challenging case, exceeds 0.2 MAE in only one case. Overall, MAE on statistical input ranges between 0.12 and 0.21 (0.14 on average) aIPC.
\subsection{Combined Agon Prediction Accuracy}
Figure \ref{fig:dnsr_impact} presents a denoiser's effect in prediction accuracy, for the two classifier architectures we study and between multilogo and oracular datasets. Each bar cluster includes three data points: \textit{NONE} corresponds to a backend-only Agon, \textit{SIMPLE} and \textit{CNN} correspond to the two denoiser architectures. For each case, the classifier's architecture is displayed under each group of benchmark multicore systems.
We observe that in almost all cases, the addition of a denoiser can improve the prediction accuracy compared to a ``vanilla'' backend-only Agon. In some cases, the \textit{CNN-CNN} Agon neural network becomes too large for the problem and training halts (weights become NaN or zero), resulting in the few cases of chance-level prediction accuracy (16-20\%) seen in the Figure.

Interestingly, the addition of a front-end module improves prediction accuracy even in oracularly trained, oracularly-fed classifier models. Our results verify the expectation of a generative (rather than denoising) front-end. Agon's front-end manipulates the (oracular in this case) input in such a way as to improve the linked classifier's accuracy. We observe the same behavior in the majority of cases of the oracular part of our results.

Our observations using the oracular dataset present some insights as to the behavior of Agon's two parts. On the other hand, results on the more realistic multilogo dataset present the real-world impact of our proposed architecture.  The addition of a simple denoiser is observed to be beneficial in many cases of our experiment. Paired with a deep classifier, a simple denoiser boosts prediction accuracy by 3.6-6.3\%. Replacing the simple denoiser with the much more complex convolutional architecture (still using a deep classifier) further increases accuracy to 4.5-11\% with the exception of the failed case.  Pairing a denoiser with a convolutional classification backend presents new behavior. First, adding a simple denoiser is no longer beneficial for prediction accuracy. The vanilla classifier's accuracy is negatively impacted by 0.7-6\% (the H8 case is improved by 0.7\%). Our results hint that our convolutional classifier is already complex enough to handle error at its inputs. It also appears that even though the CNN classifier can render a simple denoiser unnecessary, it can still (slightly) benefit from a more complex convolutional denoiser (0.5-2.6\% accuracy improvement). 

\subsection{Agon Performance}

We finally present a performance comparison of Agon in Figure \ref{fig:agon_perf}. Specifically, we compare the six Agon configurations we propose in this work, against six ``always'' scheduling strategies. We refer to a strategy as ``always'', when there is no competition between scheduling algorithms. For example, the always-RAND bar ($4^{th}$ bar from the left in each group) represents a system equipped with only a random scheduler. The first bar in each group (labeled ORACULAR) represents the best case scenario, where we oracularly pick the optimal scheduler for each problem instance. Agon configurations are labeled with \textit{[frontend]\_[backend]} keywords.  For each system, we run our 20K problem instances through each of the 13 strategies and report average system performance, normalized to ORACULAR performance. We only present results using our realistic multilogo dataset for this experiment.

Starting from the E8 system, Agon boosts performance by $\sim4\%$ compared to the most efficient always strategy (always-hungarian). Specifically, the best Agon configuration performs within 5\% of an oracular selector. The two cnn\_* Agon configurations are outperformed by most always strategies for this system, however in the remaining 3 systems, all versions of Agon outperform the best always strategy.
In very difficult (H8) systems, we see the always-HUN and always-SHUN4 (Serial Hungarian, pod size=4) cases performing much better than the rest of the always strategies. All versions of Agon outperform the best always strategy, with the \textit{none\_cnn} configuration reporting 99\% of the oracular performance (compared to 94\% for always-HUN).

Our results on the LP8 system show that always-HUN is comparable in performance to the best Agon configuration. The lowest-performing 8 core system consists of a combination of weak cores and, as a result, the majority of schedules for each problem instance will lead to roughly the same low performance. Consequently, focusing on selecting the best scheduler for each case appears to be unnecessary, even though we still measure Agon to perform within 1\% of oracle.

In contrast to LP8 however, HP8 presents a different image. First, we observe Hungarian strategies failing. This is due to the increased cost of each cycle spent in scheduling, as this system can get the maximum amount of work done in one clock cycle compared to all other systems in our study. This is also the reason we observe the low-overhead schedulers performing well. Agon (cnn\_cnn) scores 99.1\% of oracular performance (97.6-99.1\% for all Agon flavors). In comparison, the best always strategy (always-greedy) scores 95.1\%. 

In conclusion, our results reveal that a competitive scheduling approach like Agon is more beneficial when used with difficult-to-schedule systems such as H8, and with high-performance systems such as HP8.



\section{Conclusion}
\label{sec:conclusion}
This paper presents Agon, a neural network-based competitive scheduler, which successfully selects the most efficient scheduler from a list of six candidates, according to the different instances of the scheduling problem it is faced with. Agon includes a novel denoising autoencoder front-end capable of minimizing the prediction error introduced at its input by the system's performance prediction module. Agon outperforms single-scheduler approaches in all our benchmark systems, while achieving performance within 1\% of an oracular scheduler selector.


\bibliographystyle{IEEEtran}
\bibliography{references}

\end{document}